\documentclass[useAMS,usenatbib]{mn2e}
\usepackage{graphicx,psfig}
\usepackage{natbib}
\usepackage{txfonts}
\newcommand\apj{ApJ}

\newcommand\alp{$\alpha$ Centauri}

\def\be{\begin{equation}}
\def\ee{\end{equation}}

\title{Planet formation in Alpha Centauri A revisited:
not so accretion-friendly after all}
\author[P. Th\'ebault, F. Marzari, H. Scholl]
{P. Th\'ebault$^{1}$\thanks{E-mail:philippe.thebault@obspm.fr},
F. Marzari$^{2}$, H.Scholl$^{3}$\\
$^{1}$Stockholm Observatory, Albanova Universitetcentrum,
SE-10691 Stockholm, Sweden, and\\ 
Observatoire de Paris, Section de Meudon, F-92195 Meudon Principal Cedex,
France\\
$^{2}$Department of Physics, University of Padova, Via Marzolo 8,
35131 Padova, Italy\\
$^{3}$Laboratoire Cassiop\'ee, Université de Nice Sophia Antipolis, CNRS, Observatoire de la C\^ote d'Azur, B.P. 4229,
F-06304 Nice, France
}

\begin{document}

\date{Draft version \today}

\maketitle

\begin{abstract}

We numerically explore planet formation around $\alpha$ Centauri A 
by focusing on the crucial planetesimals-to-embryos
phase. Our approach is significantly improved with respect to the earlier
work of \cite{mascho00}, since our deterministic N-body code
computing the relative velocities between test planetesimals
handles bodies with different size. Due to this step up, 
we can derive the accretion vs. fragmentation trend of a
planetesimal population having any given size distribution. 
This is a critical aspect of planet formation in binaries since the 
pericenter alignment of planetesimal orbits 
due to the gravitational perturbations of the 
companion star and to gas friction strongly depends on size.  
Contrary to \cite{mascho00}, we find that, for the nominal case
of a Minimum Mass Solar Nebula
gas disc, the region beyond $\sim 0.5\,$AU from the primary is
strongly hostile to planetesimal accretion. In this area, impact
velocities between different-size bodies are increased, by
the differential orbital phasing, to values too high to allow mutual accretion.
For any realistic size distribution
for the planetesimal population, this accretion-inhibiting effect is the
dominant collision outcome and the accretion process is halted.
Results are relatively robust with respect to the profile and density
of the gas disc. Except for an unrealistic almost gas-free case,
the inner "accretion-safe" area never extends beyond 0.75AU.
We conclude that planet formation is very difficult in the terrestrial region
around $\alpha$ Centauri A, unless it started from fast-formed
very large ($>$30km) planetesimals. Notwithstanding these unlikely initial
conditions, the only possible explanation for the presence of
planets around 1 AU from the star
would be the hypothetical outward migration of
planets formed closer to the star or a different orbital configuration in
the binary's early history. Our conclusions differ from those
of several studies focusing on the later embryos-to-planets stage, 
confirming that the planetesimals-to-embryos phase is more affected
by binary perturbations.
\end{abstract}

\begin{keywords}
planetary systems: formation -- stars: individual: \alp
-- planets and satellites: formation.
\end{keywords}

\section{Introduction} \label{sec:intro}

A majority of stars are members of binary or multiple systems
\citep[e.g.][]{duq91},
making the presence and frequency of planets in such systems,
in particular terrestrial planets, a very relevant issue
in our ongoing search for other worlds. 
This issue has been made all the more timely by
the discovery of more than 40 extrasolar planets in
multiple systems, amounting to $\sim 20\%$ of all
detected exoplanets \citep{ragha06,desi07}.
Most of these planets are located in
very wide binaries ($>100$AU) where the companion star's influence
at the planet's location is probably weak
\citep[except for possible high inclination effects between the planet's
and the binary's orbits, see e.g.][]{takeda08}.
However, a handfull of planets, like Gliese 86, HD41004 or
$\gamma$-Cephei, inhabit binaries with
separations $\sim$20 AU, for which the binarity of the system
probably cannot be ignored. 

The long term stability of planetary orbits in binaries
has been explored in several studies
\citep[e.g.][]{holw99,dav03,mud06}, 
aimed at estimating $a_\mathrm{crit}$, the distance from the primary
beyond which orbits become unstable.
This critical distance depends on several parameters,
mainly the binary's semi-major axis $a_b$ and
eccentricity $e_b$ as well as the mass ratio
$\mu=m_2/(m_1+m_2)$. However, one can use as a rule of thumb
that, except for extremely eccentric systems, $a_\mathrm{crit}$
is in the 0.1-0.4$a_b$ range \citep[see for example the
empirical formulae derived by][]{holw99}. As a consequence,
the terrestrial planet region around 1\,AU is almost always "safe"
for binaries of separations $>10$AU.

The way the planet \emph{formation} process is affected by stellar binarity
is a more difficult issue. This process is indeed  a
complex succession of different stages \footnote{we consider here
the  widely accepted "core-accretion" scenario} \citep[e.g][]{lis93},
each of which could react in a different way to the
companion's star perturbations.

The last stage of planet formation, leading from planetary embryos
to planets, has been investigated in several papers
\citep[e.g][]{barb02,quin02,quin07,hag07}. These studies have revealed
that the ($a_b,e_b,\mu$) parameter space for which this
stage can proceed unimpeded is slightly, but not too much more limited
than that for orbital stability. As an illustration, the numerical exploration
of \citet{quin07} showed that planetary embryos can mutually accrete
at 1AU from the primary if the companion's periastron $q_2$ is
$\geq\,5\,$AU.

In contrast, the stage preceding this final step,
i.e. the mutual accretion of kilometre-sized planetesimals leading to the
embryos themselves, is much more sensitive to binarity effects.
A series of numerical studies \citep{mascho00,theb04,theb06,paard08}
have indeed shown that perturbations by the companion star can stir up
encounter velocities $\langle \Delta v \rangle$ to values too high
for small planetesimals to have accreting impacts. Such accretion-inhibiting
configurations are obtained for a wider range
of binary parameters than for the later embryo-to-planet phase. 
As an example, \citet{theb06} (hereafter TMS06) have shown that planetesimal
accretion in the 1\,AU region can be perturbed in binaries with separations
as large as 40 AU.
These results are not surprising in themselves, and this for two reasons.
1) The growth mode in the planetesimal-accumulation phase is the
so-called runaway growth, whose efficiency is very sensitive to increases
of the encounter velocities (which reduce the gravitational focusing factor,
e.g. TMS06). 2) It takes a much smaller
perturbation to stop accretion of a 1km body than of a lunar-sized embryo:
$\langle \Delta v \rangle>$10m.s$^{-1}$ is typically enough to achieve the
former, while large embryos will hardly feel any difference in
their accretion rate for such small velocity disturbances.
A crucial mechanism driving the $\langle \Delta v \rangle$ evolution
of the planetesimals
is friction due to the primordial gas remaining in the protoplanetary disc.
In a circumprimary disc, its main effect is to induce a strong phasing of
neighboroughing planetesimal orbits, which cancels out the large
orbital parameters oscillations around the forced eccentricity $e_f$
induced by the companion's perturbations
\citep{mascho00}. A crucial point is that this effect is \emph{size dependent},
so that while $\langle \Delta v \rangle$ between equal-sized objects are
reduced, impact velocities between objects of \emph{different sizes} are
increased. TMS06 have shown that the differential
phasing effect appears already for small size differences,
suggesting that this effect may dominate in "real" planetesimal systems.
However, the exact balance between this accretion-inhibiting effect and
the accretion-friendly effect on equal-sized
objects depends on the size distribution in the initial planetesimal population,
a parameter which could not be thoroughly investigated in the global
investigation of TMS06.

\subsection{planet formation in \alp\ A}

We follow here the approach of TMS06 to
investigate in detail one specific system: \alp\ A. Focusing on one
single system allows to take the TMS06 model a step further 
and explore realistic size distributions
of the planetesimal population as well as crucial parameters
such as gas disc density and radial profile.

At a distance of 1.33\,pc, the $\alpha$ Cen AB binary system, with a
possible third distant companion M--star Proxima
\citep{wert06}, is our
closest neighbour in space. \alp\ A and B are G2V 
and K1V stars, of masses $M_A = 1.1 M_{\odot}$ and 
$M_B = 0.93 M_{\odot}$ respectively. The binary's semi-major axis
is $a_b=23.4\,$AU and its eccentricity $e_b=0.52$ \citep{pourb02}.
\citet{holw97} have shown that the region within $\sim 3\,$AU from
the primary can harbour stable planetary orbits. However,
the search for planetary companion has so far been unsuccessful. 
The radial velocity analysis of \citet{endl01} has shown that the
upper limit for a planet (on a circular orbit) is $2.5\,M_{Jupiter}$
around $\alpha$ Cen A, and this at any radius from the star.
This leaves an open possibility for the presence of terrestrial
planets in the crucial $\sim 1\,$AU region.

Planet formation around $\alpha$ Cen A has so far been specifically
investigated in 4 studies. \citet{barb02} and \citet{quin02,quin07}
have focused on the late
embryos-to-planets stage and have found that it can proceed
unimpeded in a wide inner region ($\leq 2.5\,$AU).
For the planetesimal-accretion phase, \citet{mascho00} have
found that orbital phasing induced by gas drag allowed km-sized
bodies to accrete in the $r\leq2\,$AU region. However, 
this study only considered impacts between equal-sized objects,
thus implicitly overlooking the accretion-inhibiting
differential phasing effect.
We reinvestigate here this crucial phase,
exploring in detail gas drag effects on
realistic size distributions for the planetesimal population.

\section{Model} \label{sec:model}

\subsection{algorithm} \label{sec:algo}

We use the same deterministic N-body code as in TMS06, which follows the
evolution, in 3D, of a swarm of test particles submitted to the gravitational
influence of both stars and to gaseous friction. This algorithm tracks
all mutual encounters within the swarm and allows to derive precise
statistics of the evolution of the $\langle \Delta v \rangle$ distribution.
Gas drag is modelled by assuming a non evolving
axisymmetric gas disc, where the only free parameter is $\rho_{g}(r)$,
the radial profile of the gas density distribution. The drag force
then reads
\begin{eqnarray}  
        \vec{F} & = & - \frac{3 \rho_{\rm g} C_d} {8 \rho_{{\rm pl}} s_{}} v \vec{v}  , 
\label{drag}
\end{eqnarray}  
where $\vec{v}$ is the velocity  
of the particle with respect to the gas and $v$ the velocity modulus.  
$\rho_{{\rm pl}}$ and $s$ are the  
planetesimal density and physical radius, respectively. $C_d$ is a  
dimensionless coefficient $\simeq 0.4$ for spherical bodies.
Note that the planetesimal's
physical size $s$ considered for the drag computation cannot
be taken for the close encounter search routine, because 
the collision statistics would be much too poor with the limited number
of particles (typically $10^{4}$)
inherent to N-body codes.
We thus have to rely on the usual prescription of assigning an inflated
radius $s_{num}$ to the particles for the encounter search routine
\citep{thebra98,char01,lith07}. 
This procedure is valid as long as it does not
introduce a bias, in the form of an artificial shear component,
in the $\langle \Delta v \rangle$ statistics.
We take here $s_{num}=10^{-5}\,$AU, a value which gives a precision
of the order of 1\,m.s$^{-1}$ in the encounter velocity estimate \footnote{to
a first order, the error in the $\langle \Delta v \rangle$ estimate
is of the order of $v_{Kep}(r) \times s_{num}/r$, where $r$
is the radial distance to the star and $v_{Kep}$ the orbital velocity at distance
$r$}.

\subsection{gas disc prescription} \label{sec:gas}

The axisymmetric prescription is a simplification of the
real behaviour of the gas disc, which should also
react to the companion star's perturbations and departs from
perfectly circular streamlines \citep{arty94}. To realistically model
the evolution of both the planetesimal and gaseous components, using
a coupled N-body and hydro approach, is an arduous task.
First attempts at addressing this issue have been done by
\citet{ciecie07} for a simplified circular-binary case and by
\citet{paard08} for eccentric binaries. 
\citet{paard08} have shown that the gas disc becomes eccentric under
the companion's perturbations but that
the amplitude of this eccentricity $e_g$ depends on the choice for
the numerical wave damping procedure (the flux-limiter prescription).
There are two possible modes: a "quiet" low-$e_g$ case for
first order flux limiters and an "excited" high-$e_g$ case
for second order ones. In the "quiet" regime, the $\langle \Delta v \rangle$
statistics for planetesimals is relatively close to the simplified
axisymmetric case. In fact for a binary of 20\,AU separation, results between
the two cases were indistinguishable. In the "excited" regime, encounter
velocities are on average increased by a factor of $\sim\,2$, with respect
to the circular case, for a 20\,AU separation binary.
It is important to point out that, 
regardless of the details of the wave-damping modelling procedure,
the $\langle \Delta v \rangle$ are either equal to or higher than
their values in the circular disc case
\citep[for a more detailed discussion, see][]{paard08}.

Because of these uncertainties regarding the "real" behaviour of
the gas disc, we chose to keep here the simplified axisymmetric gas
disc approach. This choice is also motivated by the fact that these 
full planetesimal+gas simulations are still very CPU expensive and are thus
not suited for detailed parameter-space explorations as the one which is the purpose
of this study. However, in the light of the conclusions of \citet{paard08},
we think that our results should not be too different (within a factor 2 in
terms of encounter velocity values) from those expected with a full
N-body + Hydro modelling. 
In any case, a crucial point is that
our results should be regarded as giving a lower
boundary for the $\langle \Delta v \rangle$ estimates.

\subsection{set up} \label{sec:setup}

We consider for each run a number $N = 10^{4}$ of test particles, each
having the same numerical size (inflated radius) 
$s_{num}=10^{-5}$\,AU and physical
sizes (for the gas drag computation)
taken at random between two boundaries $s_{min}$ and $s_{max}$.
For our nominal case, we consider planetesimals in the 
$s_{min}=1\,$km to $s_{max}=10\,$km range. We also consider an alternative
"small planetesimals" case with $s_{min}=0.1\,$km and $s_{max}=1\,$km
and a "large planetesimals" case with $s_{min}=5\,$km and $s_{max}=50\,$km
(see Sec.\ref{sec:dfplan}).
The test particles are initially distributed in an approximate
terrestrial planet region, between $r_{min}=0.3\,$AU and $r_{min}=1.5\,$AU.
They are initially placed on almost circular orbits
($\langle e \rangle_{0}\leq 10^{-5}$), so that initial encounter velocities are
low enough to be in the accretion-friendly range.
This choice might not correspond
to a physically realistic case, but TMS06 have shown that the initial orbital
conditions are very quickly relaxed and that the system tends towards the same
steady-state regardless of the initial $\langle e \rangle$ and
$\langle i \rangle$.

For the gas density, we consider as a nominal case
the canonical Minimum Mass Solar Nebulae
(MMSN) radial profile derived by \citet{haya81}, with
$\rho_{\rm g}=\rho_{\rm g0}(r/$1AU$)^{-2.75}$ and
$\rho_{\rm g0}=1.4\times10^{-9}$g.cm$^{-3}$, but explore different 
radial profiles and $\rho_{\rm g0}$ values (see Sec.\ref{sec:gdisc}).

The final timescale for each run is taken to be $t_{\rm fin}=10^{4}$yrs, i.e.
the typical timescale for the runaway accretion of kilometre-sized
planetesimals \citep[e.g.][]{lis93}. However, this choice is not
too stringent, since in most cases a steady-state $\langle \Delta v \rangle$
distribution is reached in less than a few $10^{3}$yrs.

All characteristics of the nominal case set-up are summarised in Tab.\ref{setup}

\begin{table}
\begin{center}
\caption[]{Nominal case setup. The fields marked by a $\surd$
are explored as free parameters in the simulations. See text for
details on each parameter.}
\label{setup}
\begin{tabular}{lll}
\hline
& $r_{\rm min}$,$r_{\rm max}$ & 0.3-1.5\,AU\\
& Number of test particles & $10^{4}$\\
$\surd$ & Physical sizes $s_{\rm min}$,$s_{\rm max}$ & 1-10\,km\\
& Inflated radius (collision search) & $10^{-5}$AU\\
$\surd$ & Gas disc radial profile & $\rho_{\rm g}=\rho_{\rm g0}(r/$1AU$)^{-2.75}$\\
$\surd$ & $\rho_{\rm g0}$ & $1.4\times 10^{-9}$g.cm$^{-3}$\\
& $t_{\rm final}$ & $10^{4}$yrs\\
\hline
\end{tabular}
\end{center}
\end{table}

\subsection{$\langle \Delta v \rangle$ analysis} \label{sec:vel}

The distribution of encounter velocities is the main output we are
interested in. For each mutual impact found by the close-encounter
search routine, we record $\Delta v_{\rm s_1,s_2}$,
$s_1$ and $s_2$ being the (physical) sizes of the impacting bodies.
The encounter velocity distributions are then classified in 2-D matrixes
giving $\Delta v$ averaged over $[s_1,s_1+ds]$ and $[s_2,s_2+ds]$
size intervals in the $[s_{min},s_{max}]$ range.

The crucial issue is of course to interpret these velocities in terms
of accreting or eroding impacts. For an impacting pair of
sizes $s_1$ and $s_2$, the limit between eroding and accreting
impacts can be defined by the threshold velocity $v*_{(s_1,s_2)}$.
This velocity depends on several parameters obtained from
energy scaling considerations. The main one is the threshold energy
for catastrophic fragmentation $Q*$ \footnote{although it is important
to stress that cratering (i.e., non fragmenting) impacts play a
crucial role in defining the accretion/erosion limit, the parameters
for cratering, mainly the excavation coefficient, follow (to a first order)
similar energy scaling laws as $Q*$}, for which various and
often conflicting estimates can be found in the literature \citep[see][ for
a thorough discussion on the subject]{theb07}.
We follow the same conservative approach as in TMS06 and consider 2 boundary
estimates for the threshold velocity, $v*_{(s_1,s_2){\rm low}}$  and
$v*_{(s_1,s_2){\rm high}}$, derived using a weakly-resistant and
strongly-resistant $Q*$ prescription respectively (see TMS06 for more detail).
For velocities in between these 2 values, we do not draw any conclusions
regarding the accretion vs. erosion balance.

Note that, despite their diversity, all prescriptions for
$Q*$ agree that the size
ranges considered here, i.e. 0.1 to 50\,km, correspond to
the so-called gravitational regime, where bodies' resistance is
determined by gravitational binding forces. This means that $Q*$ is
an increasing function of size. This point is crucial for
the results of Sec.\ref{sec:dfplan}.

\section{Results}

\subsection{$\langle \Delta v \rangle$ distribution, erosion threshold}\label{sec:dvd}

\begin{figure}
\includegraphics[width=\columnwidth]{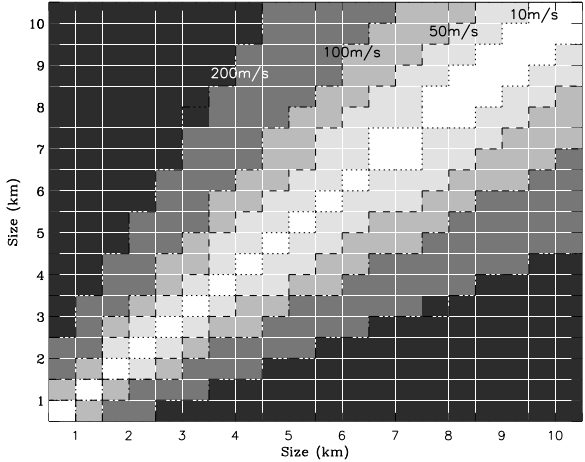}
\caption[]{Mean encounter velocities $\langle \Delta v_{\rm s_1,s_2}\rangle$,
averaged in the 0.9-1.1\,AU region,
for all impacting pairs of sizes $s_1$ and $s_2$ in the $[1,10$km$]$ range.
For the sake of clarity, 4 iso-velocity lines are represented.
}
\label{velnom}
\end{figure}

\begin{figure}
\includegraphics[width=\columnwidth]{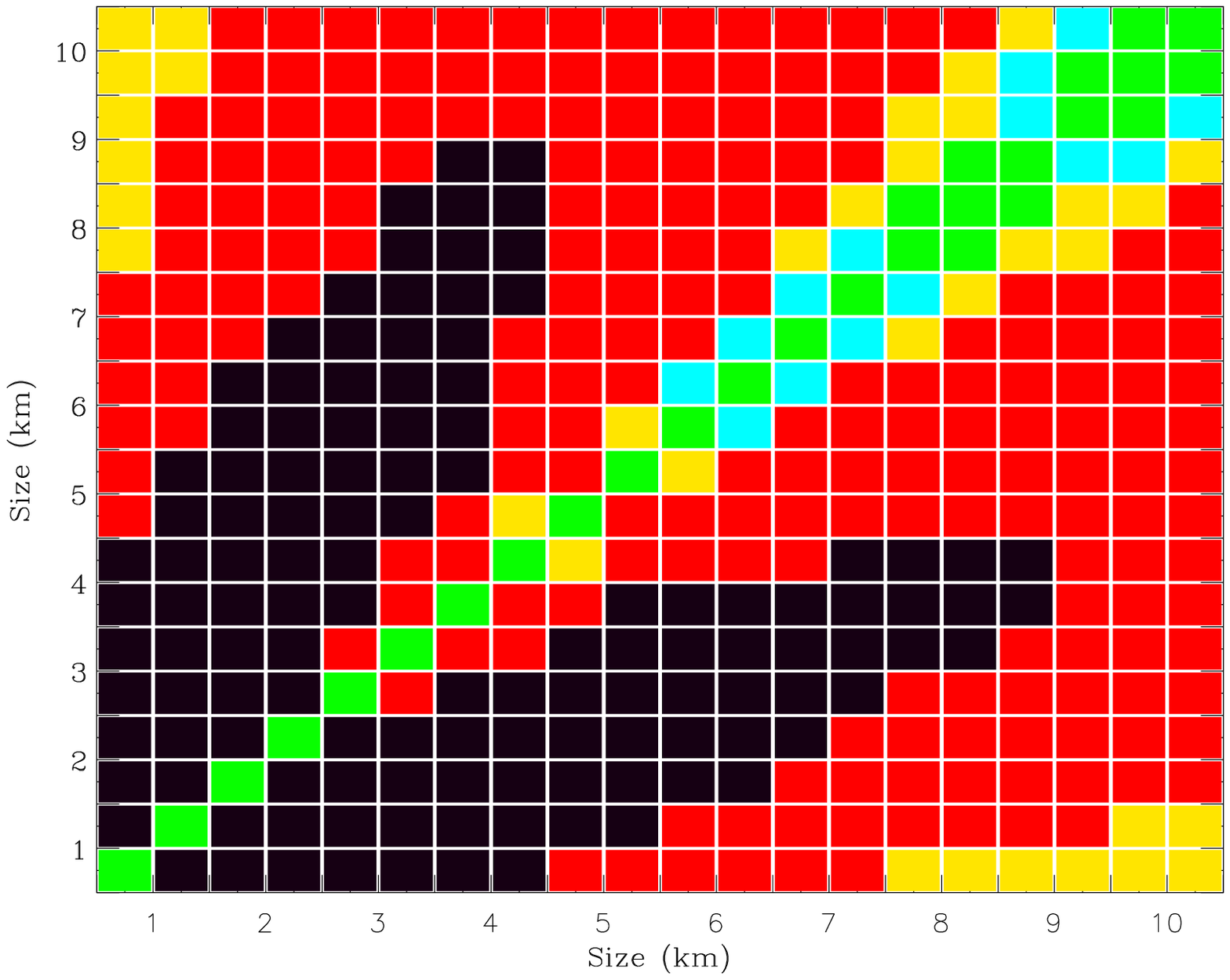}
\caption[]{Same results as in Fig.\ref{velnom}, but to each 
$\langle \Delta v_{\rm s_1,s_2}\rangle$ value is associated a colour scale,
depending on how this value translates into accreting or erosive impacts
for each impacting pair of sizes $s_1$ and $s_2$:

-{\bf green}: no $\langle \Delta v \rangle$ variation observed, region of \emph{ unimpeded accretion}

-{\bf blue}: encounter velocity increase but 
$\langle \Delta v \rangle \leq v*_{\rm low}$, \emph{perturbed accretion with
slow-down of the runaway growth}

-{\bf yellow}: $ v*_{\rm low} \leq \langle \Delta v \rangle
\leq v*_{\rm high}$, \emph{ uncertain accretion vs. erosion balance}

-{\bf red}: $\langle \Delta v \rangle \geq v*_{\rm high}$, \emph{ erosion}

-{\bf black}: $\langle \Delta v \rangle \geq 2\times v*_{\rm high}$, \emph{ erosion}

}
\label{accnom}
\end{figure}

Fig.\ref{velnom} shows the $\langle \Delta v_{\rm s_1,s_2}\rangle $ 
distribution, at 1\,AU from the primary and for bodies in
the $[1,10$km$]$ range, once a steady-state
is reached, i.e. after $\simeq 3\times 10^{3}\,$yrs.
The sharp size dependence of encounter velocities is here clearly
visible. Low $\langle \Delta v \rangle$ are restricted
to a narrow stripe along the $s_1=s_2$ diagonal, corresponding to
the domain where gas drag aligns orbits towards the same value.
However, for any small departure from the exact $s_1=s_2$ condition,
encounter velocities dramatically increase. We get
$\langle \Delta v_{\rm s1,s2}\rangle \geq 50$m.s$^{-1}$ in almost
the whole $s_1 \neq s_2$ region. This illustrates very clearly
the strength of the differential orbital phasing mechanism.

Fig.\ref{accnom} shows how these velocities can be interpreted
in terms of accreting/erosive impacts. The
only region where encounter velocities are low enough
for the accretion process to proceed unimpeded (in green) is the
narrow $s_1=s_2$ diagonal. This diagonal is surrounded,
for objects $>$5\,km, by an equally narrow band (in light blue)
where $\langle \Delta v\rangle$ allow accretion
($\langle \Delta v\rangle_{\rm s_1,s_2} \leq v*_{(r_1,r_2){\rm low}}$)
but are nevertheless increased with respect to their initial
unperturbed value. This means that the accretion rate is slowed
down, compared to the unperturbed case, since the runaway growth mode's efficiency 
decreases with increasing encounter velocities.
Except for a small region (in yellow) of uncertain accretion/erosion
balance where
$v*_{(s_1,s_2){\rm low}} \leq \langle \Delta v\rangle_{\rm s_1,s_2} \leq
v*_{(s_1,s_2){\rm high}}$, all other ($s_1,s_2$) encounters result
in eroding impacts. In these red and black areas, which cover
most of the graph, velocities exceed $v*_{(s_1,s_2){\rm high}}$,
our most conservative estimate for the threshold velocity .

\subsection{accreting vs eroding impacts: size distribution} \label{sec:accero}

\begin{figure}
\includegraphics[width=\columnwidth]{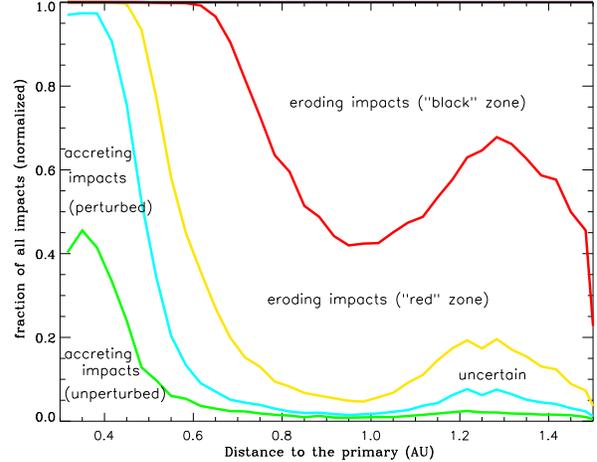}
\caption[]{Relative importance of different types of collision outcomes
(corresponding to the different colour categories of Fig.\ref{accnom}),
as a function of distance to the primary star. All close encounters
between all bodies in the $[s_{min}$,$s_{max}]$ size range are taken into account,
with each close encounter weighted assuming that the
size distribution for the planetesimals follows a Maxwellian 
distribution centered
on 5km (see text for details).
}
\label{accstat}
\end{figure}

Of course, Fig.\ref{accnom} cannot in itself give the relative
importance of the accreting-friendly vs. accretion-hostile areas.
To do so, the real \emph{size distribution} of the planetesimals should be
known, whereas in the runs all sizes were chosen at random between $s_{min}$
and $s_{max}$. The size distribution of a typical
initial planetesimal population is a complex issue,
which is directly linked to the very difficult (and yet unresolved)
problem of the mechanisms for planetesimal formation. This problem
exceeds by far the scope of the present paper (see, however, the discussion in
TMS06 and in the last Section of the present paper). 
In line with our conservative approach,
we shall here assume a fiducial size distribution which is {\it a priori}
favourable to same-size impacts (that is, to
accretion), i.e. a Maxwellian $f_{mx}(s)$ peaking at a median value $s_{med}=5\,$km.
\footnote{Some planetesimal-accumulation particle-in-a-box
models make the assumption of a single-size initial population
\citep[e.g.][]{weiden97}, but this is of course
a numerically convenient simplification which cannot correspond
to a "real" population which necessarily has a spread in sizes}.
We then weight each ($s_1$,$s_2$) impact by
$f_i=f_{mx}(s_1)\times f_{mx}(s_2)$, the (normalised) probability of such
an impact with a $f_{mx}(s)$ size distribution. Depending on
$\langle \Delta v\rangle_{\rm s_1,s_2}$, all impacts are then
classified into one of the 5 collision-outcome categories defined 
earlier (see Fig.\ref{accnom}), where their contribution is
weighed by $f_i$. Comparing the total $\Sigma f_i$ for each
category then gives the relative importance of each of these
possible collision outcomes.

As can be seen in Fig.\ref{accstat}, at 1\,AU less than 2\% of all impacts
result in accretion (a value which only increases to $\sim 5\%$ when including
the "uncertain-outcome" category), while 95\% should lead to net erosion of
the impacting bodies. The dynamical environment is thus globally very hostile
to accretion. Fig.\ref{accstat} also shows that this
accretion-hostile environment extends over most of the system.
Only the innermost $r\leq 0.5$\,AU region has a favourable
accretion vs. erosion balance. This accretion-friendly inner zone
is mostly due to weaker gravitational
perturbations by the companion star.

Between 1.2 and 1.4\,AU, a slight increase in
the accreting/eroding impacts ratio is observed, which is mostly due to
weaker gas drag-induced differential phasing. However, even there
the fraction of impacts with
$\langle \Delta v \rangle \leq v*_{\rm low}$ never exceeds 8\%.
Moreover, beyond 1.4\,AU, accreting impacts
quickly decrease to 0 again. The reason for high $\langle \Delta v \rangle$
in these outermost regions is no longer gas friction
but pure gravitational effects. Indeed, although
in the absence of gas drag the companion star forces secular orbital
oscillations around $e_f$ which should in principle
not induce high $\langle \Delta v \rangle$
because they preserve the same phasing for all bodies
(regardless of their size), these oscillations get narrower
with time and at some point orbital crossing occurs, leading to
an abrupt relative velocity increase (for a more detailed description of this
effect, see section 2 of TMS06). Strong gas drag can prevent
this crossing from happening, by cancelling out the oscillations
and inducing it's own size-dependent phasing, but in the outer regions it
is too weak to do so.
In other words, planetesimals are here caught between a rock and a
hard place:  \begin{itemize}
\item{In the outer regions, pure gravitational perturbations
lead to orbital crossing and high $\langle \Delta v \rangle$}
\item{In the inner regions, gas drag can prevent orbital crossing but induces
size-dependent orbital phasing which also leads to high $\langle \Delta v \rangle$}
\end{itemize}

\begin{figure}
\includegraphics[width=\columnwidth]{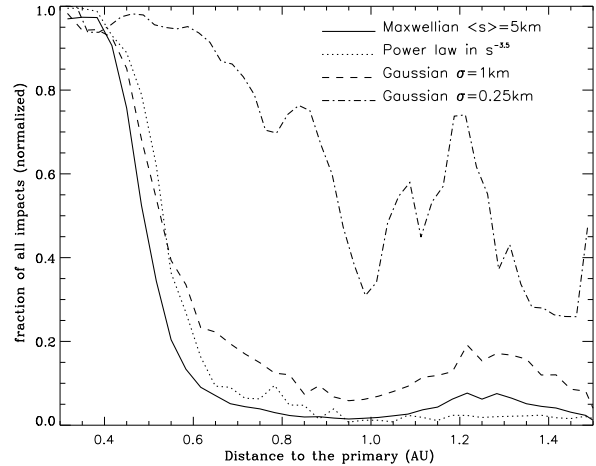}
\caption[]{Fraction of accreting impacts (limit between the "blue" and "yellow"
zones) for different prescriptions for the size distribution. The noisiness of
the curve corresponding to the extremely narrow-Gaussian is due to the fact
that in this case only impacts for a very narrow $s_1\sim s_2$ region have
non-negligible weigh in the global statistics.

}
\label{szdist}
\end{figure}

We have tested the robustness of our results with respect to 
different choices
of the size distribution. Collisional-"equilibrium" power laws 
with $dN \propto s^{-3.5}ds$ and Gaussians
centred on $s_{med}$ of various widths have been considered. 
Although some differences were
observed with respect to our nominal case, they remain relatively marginal and
our main conclusions seem to hold regardless of the
size distribution (see Fig.\ref{szdist}).
This is true in particular for the region around 
1\,AU which is always strongly hostile to accretion: the fraction of
accreting impacts never exceeds 10\% for all tested size distributions.
The only case giving an accretion-friendly environment is
an almost Dirac-like Gaussian of variance $(0.25\,{\rm km})^{2}$, but this size
distribution is probably highly unrealistic.
Another important robust result is that there is always an inner region
having low $\langle \Delta v\rangle$ enabling accretion. The outer edge of
this inner zone varies slightly with the size distribution prescription,
but is always located in the $0.5-0.6$\,AU range.

\subsection{Different gas disc profiles} \label{sec:gdisc}

\begin{figure}
\includegraphics[width=\columnwidth]{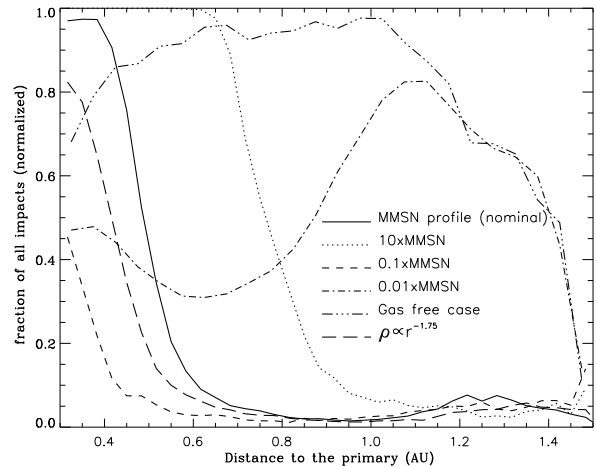}
\caption[]{Fraction of accreting impacts (limit between the "blue" and "yellow"
zones) for different prescriptions for the gas disc (see text for details).
}
\label{gsdist}
\end{figure}

For our nominal case, we assumed a standard MMSN gas disc, but 
other gas disc models have been explored (Fig.\ref{gsdist}).

We first consider the same $\rho_{\rm g}=\rho_{\rm g0}(r/$1AU$)^{-2.75}$
profile as in the nominal case but with a different reference value for $\rho_{\rm g0}$
at 1\,AU chosen around the MMSN 
$\rho_{\rm g0(ref)}=1.4\times10^{-9}$g.cm$^{-3}$ value.

As it can be seen, for larger values of $\rho_{\rm g0}$ the situation becomes
slightly more favourable to accretion, and the outer edge of the inner
accretion-friendly zone moves to $\sim 0.75\,$AU for the highest
$\rho_{\rm g0}= 10\times\rho_{\rm g0(ref)}$ explored value. This is due to the
fact that complete differential phasing, meaning that
objects of all sizes reach steady-state
eccentricity and periastron longitude $\omega$ values before $t_{\rm final}$,
is already achieved for the nominal MMSN case. Thus, when increasing the gas
density, no "better" phasing is achieved. The effect of increased $\rho_g$
is then that all steady-state $e$ and $\omega$ values globally tend
towards their asymptotic value for infinitely small sizes, i.e.
$e=0$ and $\omega=270^{o}$ \citep[see Eqs.24-29 of][]{paard08}, which
tend to reduce encounter velocities.

The effect of decreasing gas densities is more complex to interpret.
For $\rho_{\rm g0}$ just below $\rho_{\rm g0(ref)}$,
relatively good differential phasing is
maintained {\ but} all steady-state $e$ and $\omega$ values globally move away
from their asymptotic $s \rightarrow 0$ value. This results in a wider spread
of steady-state orbital parameters and thus higher $\langle \Delta v\rangle$.
Below a threshold value ($\sim 0.1\times$MMSN),
however, steady state phasing is no longer achieved
for the biggest bodies, which then still have substantial $e$ and $\omega$ 
oscillations after $t_{\rm final}$.
Since phasing is still achieved for the
smallest objects, the \emph{difference} between the orbits of small and big bodies
is still increasing. This is what is clearly observed for the
$\rho_{\rm g0}= 0.1\rho_{\rm g0(ref)}$ run. However, below a second
threshold value ($\sim 0.01\times$MMSN),
steady state phasing breaks down for \emph{all} bodies.
The system now moves towards its gas-free behaviour, with large
common orbital oscillations for all bodies, and high $\langle \Delta v\rangle$
only in the regions where oscillations are narrow enough for orbital
crossing to occur, i.e. beyond $\sim 1.3$AU (see discussion
in the previous section).

We finally consider a case with a different radial dependence for
the gas density, i.e. a flatter profile in 
$\rho_{\rm g}=\rho_{\rm g0}(r/$1AU$)^{-1.75}$ ($\rho_{\rm g0}$ being the same
as in the nominal case). The accretion/erosion balance is slightly less
favourable than in the nominal MMSN case, with the outer edge of the
inner "safe" zone at $\sim 0.45$\,AU (dashed curved in Fig.\ref{gsdist}).

However, despite all these quantitative differences, the qualitative results
remain relatively similar for all test gas disc cases, with the exception
of the probably unrealistic very-gas-poor runs. The generic characteristics
are an inner accretion-friendly
zone, whose outer edge varies from 0.35 to 0.75\,AU, followed by a vast
region where impacts are predominantly erosive for all objects in
the 1-10\,km range. More importantly, the crucial region for habitable 
planets around 1\,AU is always in
an environment hostile to accretion.

\subsection{Alternative planetesimal size-ranges} \label{sec:dfplan}

\begin{figure}
\includegraphics[width=\columnwidth]{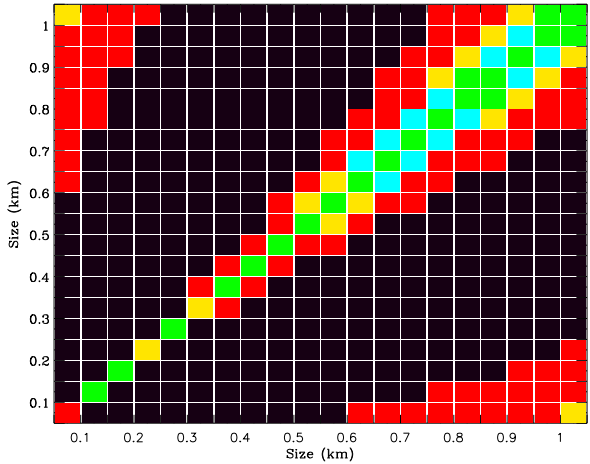}
\caption[]{Same as Fig.\ref{accnom}, but for the "small planetesimals" run.
The colour scaling for the different accretion/erosion zones is the same as 
in Fig.\ref{accnom}.
}
\label{accsmall}
\end{figure}

\begin{figure}
\includegraphics[width=\columnwidth]{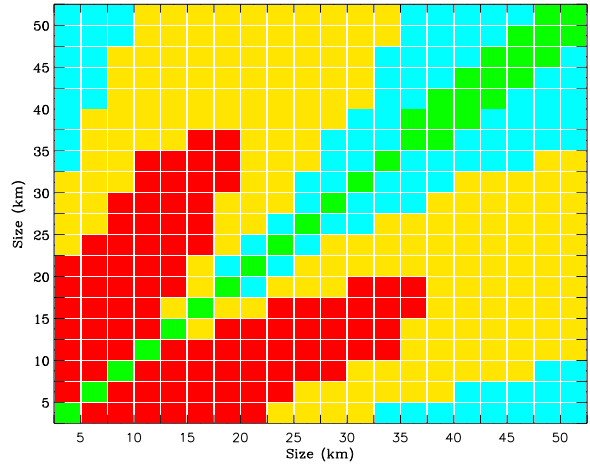}
\caption[]{Same as Fig.\ref{accnom}, but for the "large planetesimals" run.
The colour scaling for the different accretion/erosion zones is the same as 
in Fig.\ref{accnom}.
}
\label{accbig}
\end{figure}

We have so far assumed a planetesimal population in the 1-10\,km size range,
which corresponds to the most generic choice for an "initial" planetesimal
population \citep[see e.g.][]{gre78,wet89,bar93,lis93,weiden97}. However, 
because the planetesimal-formation phase is still poorly understood
(see for example \citet{you04} and \citet{cuz06} for conflicting views on
the matter), this few-km size range is far from being a value carved in marble.
Two other possible initial size ranges, with $0.1\leq s\leq1\,$km and
$5\leq s\leq50\,$km, have thus been explored in our simulations.
Note that if from a purely dynamical point of view changing the planetesimal
sizes is equivalent to changing the gas density (because of the $1/s$ dependence
of the gas drag force), it is not equivalent from the point of view
of the accretion/erosion behaviour. As an example, an impact between
a 1km and a 5 km body in a gas disc of density $\rho_g$ occurs on average
at the same velocity $\langle \Delta v \rangle_{1,2}$
as an impact between a 10km and 50km body in a $10\times\rho_g$ disc, but
the outcome of a $\langle \Delta v \rangle_{1,2}$ collision between
1km and 5km objects will not be the same as for 10km and 50km objects

For our "small" planetesimals population in the 0.1-1\,km range, 
we obtain, not surprisingly, a situation which is more hostile to
accretion (Fig.\ref{accsmall}). This is not due to higher relative velocities
than in the nominal case, since $\langle \Delta v \rangle_{1,2}$ are of the 
same order than in the 1-10\,km run. The explanation is to be found in
the fact that such small objects are much more 
affected by $\langle \Delta v \rangle$ increase, and this for 2 reasons:
1) They have lower threshold $Q*$ values, and 2) they have lower
escape velocities.

For the "large" planetesimals run with $s_{min}=5\,$km and $s_{max}=50\,$km,
on the contrary, we find an environment which is more favourable
to accretion than in the nominal case (Fig.\ref{accbig}). The
accretion-friendly zone now covers a large fraction of the ($s_1,s_2$)
phase space at 1\,AU, especially in the $\geq 30\,$km range.
As for the small planetesimals case,
this result is only marginally due to lower relative velocities,
since the distribution of $\langle \Delta v \rangle_{1,2}$ is 
here again close to that of the nominal case. What makes a difference
here is that bodies in the 10-50\,km range are much more resistant
to impacts than smaller ones. They are also big enough to reaccumulate
a substantial fraction of the mass they might lose after an impact
(a process which is taken into account in our estimate of
$v*_{(s_1,s_2){\rm low}}$ and $v*_{(s_1,s_2){\rm high}}$).

\section{Discussion and Conclusions} \label{sec:concl}

Our numerical exploration has shown that for our nominal case
set-up the $r\geq 0.5\,$AU region
around \alp\ is hostile to kilometre-sized planetesimal accretion.
The coupling of the companion star's secular perturbations
with the size-dependent phasing imposed by gas drag leads to
high $\langle \Delta v \rangle_{\rm(s_1,s_2)}$ destructive collisions for
all $s_1 \neq s_2$ pairs. 
For any realistic size distribution for the planetesimal population
(except very unlikely Dirac-like size distributions),
these erosive impacts outnumber by far the low-velocity $s_1 \sim s_2$ encounters
and are the dominant collision outcome.
Only the innermost region within $\sim 0.5\,$AU has
$\langle \Delta v \rangle$ low enough to allow planetesimal
accretion. 

Interestingly, these results are relatively robust when varying the
gas disc density. The main difference is that the outer limit
of the inner "accretion-safe" region varies between 0.35 and 0.75\,AU.
Only for very gas poor systems do we get a qualitatively different
behaviour, with the safe region extending up to $\sim 1.3\,$AU, where
a high-$\langle \Delta v \rangle$ zone due to pure
secular perturbations begins.

These results are further strengthened by the fact that 
our simplified assumption for the gas disc probably results in
a slight underestimate of $\langle \Delta v \rangle_{\rm(s_1,s_2)}$ values
(see discussion in Sec.\ref{sec:gas}). As a consequence, our approach
can be regarded as conservative regarding the extent of the
impact velocity increase which is reached, the "real" system
being probably even more accretion-hostile than in the presented results.

This hostility to kilometre-size planetesimal accretion is of
course a major threat to planet formation as a whole, of which it is a crucial
stage. However, before reaching any definitive conclusions, one has to see
if there might be some ways for planetary formation to get round this hurdle.
In the next 3 subsections we critically examine three such possibilities.

\subsection{Accretion from large initial planetesimals?}

We have seen that, apart from an unrealistic gas-free case, the only way
to get an accretion friendly environment at 1\,AU is
to assume large, $\geq 30\,$km planetesimals which can sustain
the high velocity regime without being preferentially eroded
(Fig.\ref{accbig}). This solution to the high velocity problem
raises however important issues.
One of them is how realistic it is to assume a population
of large initial planetesimals. As already pointed out, the
planetesimal formation process is still too poorly understood to
reach any definitive conclusion regarding what an initial planetesimal
population looks like, so that values
in the 30-50\,km range cannot be ruled out \emph{a priori}.
Nevertheless, one has to ask the question of what was \emph{before}
these large planetesimals.
Should these objects arise from the progressive accumulation of
grains, rocks and pebbles in a turbulent nebula \citep[as in the
standard coagulation model, e.g.][]{dul05}, then there is no clear cut
separation between the planetesimal-formation phase and the subsequent
planetesimal-to-embryo phase \footnote{In this case, the very concept of
an "initial" planetesimal population might be questioned
\citep[see for instance][]{wetina00}}. In this case, the stage
where the biggest planetesimals are $\sim$50\,km in size
should be preceded by a stage where the biggest objects are in
the 1-10\,km range, which leads us back to the accretion-inhibiting
nominal case.
The only solution would be to bypass these intermediate accretion-inhibiting
stages by directly forming large planetesimals. This could in principle
be achieved with the alternative gravitational instability scenario
\citep[e.g][]{gold73,you04},
but this scenario still has several problems to overcome \citep[see for
instance][ for a brief review]{arm07}. Moreover, it is not clear
how gravitational instabilities should proceed in the highly perturbed
environment of a binary.

It should also be noted that even \emph{if} at some point this accretion-friendly
large-planetesimals stage is reached, the way accretion will proceed from there
would be very different from the standard runaway-accretion scheme. Indeed, 
as can be clearly seen in Fig.\ref{accbig}, the ($s_1$,$s_2$) parameter
space for which accretion can proceed unimpeded is restricted to a narrow
$s_1 \sim s_2$ band. For most impacts on $s\geq 30\,$km targets,
even if relative velocities are below $v*_{(s_1,s_2){\rm low}}$
(the "blue" zone), they are nevertheless increased relative
to the unperturbed case. These increased
$\langle \Delta v \rangle$ would strongly reduce the gravitational focusing
factor which is the key process driving runaway accretion. In this case, the
growth mode could be the so-called "type II" runaway identified by
\citet{kort01}, corresponding to a slowed-down runaway growth
in a high $\langle \Delta v \rangle$ environment.

\subsection{Accretion once gas has dissipated?}

An interesting issue is what happens \emph{after} 
gas dispersal. Indeed, it is believed that primordial gas discs
dissipate, due to photo-evaporation and/or viscous accretion onto
the central star, in a few $10^{6}$ years \citep[e.g][]{alex06},
a timescale which could even be shortened in binaries because of 
the outer truncation of the protoplanetary disc. One might thus wonder
if, despite of a long accretion-hostile period while gas is present,
planetesimal accretion could start anew once the environment becomes
gas free and differential phasing effects disappear. 
However, for this to occur, there is a big difficulty to overcome.
Indeed, when the environment becomes gas free, all planetesimals are
still on the differentially-phased orbits they were forced onto by gas
friction. This means that the \emph{initial} dynamical conditions are such as
$\langle \Delta v \rangle_{\rm(s_1,s_2)}$ are high. Under the sole
influence of the companion star's secular perturbations (the only external
force left in the gas free system), these initial
orbital differences cannot be erased or damped. The size-independent orbital
phasing which will be forced by secular effects will come \emph{in addition}
to the initial orbital differences, which can in this respect be regarded
as an initial free component. This behaviour can be easily checked by
gas free test simulations. 

Moreover, even if a hypothetical additional mechanism acts to damp the initial 
differential phasing, it has to act quickly in order to lead to low
impact velocities. Indeed, if given enough time, secular perturbations
\emph{alone} will eventually induce high $\langle \Delta v \rangle$ due to orbital
crossing of neighbouring orbits. The inner limit of $\sim$1.3\,AU for such
orbital crossing given in Sec.\,\ref{sec:accero} is indeed the one reached
at $t=10^{4}yrs$, but this limit is moving inwards with time. Using 
the empirical analytical expressions derived by TMS06 (see Eq.15 of that
paper), we find that the orbital crossing "front" reaches the 0.5\,AU region
after $\sim 1.5\times 10^{5}$yrs.

As a consequence, it seems unlikely that $\langle \Delta v \rangle_{\rm(s_1,s_2)}$
can be significantly reduced in the later gas free system. Collisional
erosion, started in the gas-rich initial disc, should thus probably
continue once the gas has dissipated
\footnote{As this paper was being reviewed, a yet unpublished
study by Xie and Zhou \citep[ApJ, submitted, with a prelimimary summary
in][]{xie08} was brought to our attention. It
explores the possibility, for the specific case of $\gamma$ Cephei and
under possibly extreme assumptions for gas dispersal timescales,
that some "re-phasing" might occur {\it during} the gas dissipation phase}.

\subsection{Outward planet migration and different initial binary configuration}\label{sec:diss}

It could be argued that planets could have formed elsewhere and later
migrated to the $r\geq 0.75$AU regions (the maximum extent
of the accretion-friendly region obtained in our gas-disc profile exploration).
Early migration of the protoplanets by
interaction with the primordial gas disc might be ruled out since
such a migration \citep[be it of type I, II or III, e.g][]{papa07}
would be inwards. In the present case, the planets would thus 
have to come from further out in the disc, i.e,
from regions which are even more hostile to accretion.
Outward drift of planets could however occur for another type
of migration, the one triggered, after gas-disc dispersal, by interactions
between a multi-planets system and a massive disc of remaining planetesimals
\citep[as in the so-called Nice-model, e.g.][]{tsig05}. Netherless,
it remains to see how this scenario, which was studied for
mutually interacting giant planets at radial distances between 5 and 30\,AU,
would work in the present case, $\sim 10$ times closer to the star
and with probably much less massive planets. This important issue will
be addressed in a forthcoming paper (Scholl et al., in preparation).

An alternative possibility would be that the binary's orbital
configuration was not the same in the past as it appears today. If the
presently observed binary system was part of an unstable multiple system
(triple or more), planets might have formed when the binary had a larger
semi-major axis and it was less eccentric \citep{marbar07}.
Both these conditions could lead to a more favourable environment
for planetesimal accretion than present.
In this scenario planets should have formed prior to the onset
of the stellar chaotic phase which inevitably ends with the ejection of one
star, leaving the remaining binary system with smaller separation and higher
eccentricity. The discussion of the possible consequences of a transient
triple system stage for \alp\ A and B is complicated by the presence of
Proxima Cen, a red dwarf located roughly a fifth of a light-year from the
AB binary. A recent paper by \citet{wert06} 
argues that the binding energy of Proxima Centauri relative to
the centre of mass of the $\alpha$ Centauri binary is indeed negative,
even if the star is close to the outer border of the Hill's sphere of the
AB system in the galactic potential. This might be an indication of a
violent past of the system when the AB binary was possibly part of a
higher multiplicity unstable stellar system. This issue exceeds the scope
of the present paper and should be explored in future studies.

\subsection{Conclusion and Perspectives}

Because of the hypothetical additional effects listed in
Sec.\ref{sec:diss}, the presence of planets beyond 0.75\,AU
cannot be fully ruled out. However, we think that our results
on planet \emph{formation},
with the present binary configuration, are relatively robust.
Our main result is that it is
very difficult for $s\leq30$\,km planetesimals to have accreting
encounters beyond 0.5-0.75\,AU from the primary, which makes planet
formation very unlikely in these regions.
These results are in sharp contrast to those of \citet{mascho00}, who
found the terrestrial region aournd \alp\ A
to be favourable to planetesimal accumulation.
The main reason for this discrepancy is that \citet{mascho00} restricted their
pioneering study to collisions between \emph{equal-size} bodies, for which
gas friction indeed decreases $\langle \Delta v \rangle$ and favours
accretion, thus implicitly neglecting the dominant effect of eroding
$s_1 \neq s_2$ impacts. 
Our results are also in contrast to the conclusions of
\citet{barb02} and \citet{quin02,quin07}, who regarded planet accretion as been
possible in the inner $\leq 2.5\,$AU region. This is because these
studies focused on the later embryo-to-planet phase, thus implicitly
assuming that the preceding planetesimal-to-embryo phase
was successful. The present study shows that this is probably not the case.
This confirms that the planetesimals-to-embryos phase is more
affected by the binary environment than the last stages of
planet formation.

Due to the complexity of the coupling between secular
perturbations and gas drag, and their strong dependence on binary orbital
elements, the present results cannot be directly extrapolated to
other systems. Clearly, such detailed studies of planetesimal accumulation
in other close binary systems, either those
known to harbour planets (like $\gamma$ Cephei \footnote{$\gamma$ Cephei
has already been investigated by \citet{theb04}, but with a simplified
version of the present code, in particular with respect to the
role of planetesimal size distributions} or HD41004) or those
being potential candidates, should be undertaken in the close future.
In this respect, one object of particular interest could be $\alpha$ Cen A's
companion, Alpha Centauri B, for which \citet{guedes08} very recently explored
the possibilities for observational detection of terrestrial planets.

{}
\clearpage

\end{document}